   \documentclass[12pt,superscriptaddress, preprintnumbers,amsmath,amssymb]{revtex4}
\usepackage{graphicx} 
\usepackage{dcolumn} 
\usepackage{bm} 
\include{epsf}

\begin{document}

 \title{ Econophysics:  Comments on a few Applications, Successes, Methods,  \& Models\\}

\author{Marcel AUSLOOS$^{a,b,c}$ \\ 
$^a$eHumanities group,
Royal Netherlands Academy of Arts and Sciences, Joan Muyskenweg 25, 1096 CJ Amsterdam, The Netherlands
 \\  $^b$r\'es. Beauvallon, rue de la Belle Jardini\`ere, 483, \\
B-4031 Li\`ege, Wallonia-Brussels Federation\\ 
$^c${\it previously at} GRAPES@SUPRATECS, ULG, B5a Sart-Tilman,\\ B-4000 Li\`ege, still in Belgium, Euroland
\protect\\ e-mail address: marcel.ausloos@ulg.ac.be }

FOREWORD :
 preprint prepared for IIMK Society \& Management Review Vol 2(2), July 2013,  Special issue on "Econophysics: Perspectives and Prospects"  

\date{today}

 
 \begin{abstract}
For this special issue, the article aims   at discussing   
a few econophysics problems studied so far rather successfully.   
The following "applications"  in micro-econo-physics are considered : (i)  financial crashes; it is emphasized that  one can distinguish between endogenous  and exogenous causes; (ii) portofolio control, selection and inherent risk   measure; (iii) foreign currency exchanges, also distinguishing endogenous and exogenous money control;   (iv)  price and asset  evolution values.  It is shown  that some macro-econo-physics problem have been also tackled,  like  geographic/political constraints,   the globalization of the economy and country clustering.  Moreover,  it is daring to suggest  prospect for  studies and researches, whence presenting  some selection of  a few  interesting perspectives. 

\end{abstract}

\maketitle

\bigskip

{\bf  \hspace{65mm} If  a physicist wants (= tries!)  } 

{\bf  \hspace{50mm} to sell the solution of a  problem to an economist, }

{\bf  \hspace{60mm} he (the physicist) must pay a (the ?) price. }

\section{Introduction}\label{intro}
{\it  Does Econophysics Make Sense ?} asks Stauffer in Feb. 1999 \cite{DSmakesense?}.
 Econophysics is a recent branch of physics  that applies Statistical Physics methods to problems concerned with financial features   and with practical problems in
micro- and macro-economics. 
As early as 1693, at  the econo-statistician level, Halley had showed how detailed records of births
and deaths, in Breslau, now Wroc\l{}aw, PL, could be used to compute life expectancies and published the first detailed mathematical analysis of the valuation of annuities  \cite{Halley,CieckaHalley}.
Newton \cite{Newtontrad} and Gauss also  theorized financial speculation. Newton's financial investment  losses (in the South Sea bubble burst)  are well reported \cite{Newton,GleickNewton}.  Gauss was much more successful \cite{Gaussbio}.

 Bachelier, a mathematician,   developed a ``theory of
speculation'', in his 1900 Ph. D. thesis  \cite{Bachelier}.
Bachelier's work suggests a  practical connection between stochastic theory (random walks or Brownian motion) and financial analysis.  His  theory,  for option pricing, was received with extreme skepticism at 
the time (even Poincar\'e was rather critical about his student's
views). Forgotten for more than half a century, Bachelier's work is
nowadays considered as a milestone in Econophysics \cite{MantegnaStanleybook}.    Another milestone is Mandelbrot's 1963 observation 
of power-law scaling in commodity markets  \cite{BBM63}; see also  \cite{BBM67}.  

Of course, practical economists have been  running computer simulations of financial markets models for a long time.   However,  one can admit that models in modern quantitative financial analysis were stimulated by the work of physicists, e.g., Osborne \cite{Osborne} who in the  1950's rediscovered the Brownian motion signature in stock market dynamics. 

A  major achievement in the field was the early 1970's 
work of Black and Scholes \cite{BlackScholes} and of Merton \cite{Merton}    whose model casts the option pricing problem \cite{Mertonbook} into a diffusion type equation, most often referred to to-day as the Black-Scholes equation.   It can also  be  embedded   in a quantum physics setting, as shown by  Haven \cite{Havenoption}.  
 However, it  was to physicists to show that the Black-Scholes equation is based on erroneous hypotheses   \cite{BouchaudBlackScholes94}.

To make it short,  physicists  are now
applying concepts from Statistical Physics such as the  ideas of 
universality, scaling,  game theory, networks, and agent  interaction models to practical financial issues, such as  financial crashes, the movement of stock prices and portfolio management, 
currency exchange rate fluctuations, share price evolution, etc..  Moreover,  many recent  attempts turn toward contributions to the theory of macro-economics. Papers now being published  
span wider areas, such as consumer choice \cite{sales,music1} and  social trends \cite{QJE127.12.1altruism},  
business cycles \cite{sanglierauslooscycles} and the (structural and financial) evolution of organisations  \cite{garcia,MAarXiv1201.4841}
which  are also of interest to social scientists and managers.

Whence, it is aimed here below to discuss a few  problems, tackled in econophysics,  having in mind a readership of
students/teachers of economics \& business management level,  wishing to be aware of   questions raised by economists and somewhat understood by physicists interested in applying usual methods of investigations and modeling techniques  to such economy and finance problems. Only
a few econophysics problems studied so far  are presented below, - the  selection being   surely very biased, since focusing on the author's activities and interests. 

 Thus, after this introductory Sect. \ref{intro},  suggesting the immense work field  available for investigations, a few cases are  outlined, in Sect. \ref{Sect2}, first stressing 
 financial crashes  in Sect. \ref{ Crashessubsect}.  It is emphasized that ''external fields'' are   a major ingredient, in order to distinguish endogenous from exogenous causes, rendering the study more (or very) ''interesting'', within a mechanistic approach based on parity and symmetry of  concepts.   In Sect. \ref{Portofoliosubsect},  portofolio selection and inherent risk are  shown to have been considered. 
 
 Foreign Exchange  markets have been also  much studied, - the money volumes being very large, whence also gains and losses. Several features  are  discussed in Sect. \ref{FXsubsect}.  
  Finally, predictability is, at a very general level,  a major concern of agents, being traders, investors, managers or politicians. The point is discussed in Sect. \ref{Predictabilitysubsect}.   Finally,  a comment on price evolution is necessary; see self-citation at the beginning of the main text). it is presented in Sect. \ref{Pricesubsect} with some introduction to models on asset distributions.

 In Sect. \ref{Macroecono},   considerations more prone to macro-economy questions, like the globalization of the economy or (country) economic clusters are shown to have been considered in recent econophysics work.

A few short comments, in Sect. \ref{discuss}, are tied to daring  suggestions and   prospects for  studies and researches, whence presenting  some selection of  interesting (?) perspectives. 

A few Appendices. contain technical details which can be found  with much more  information in many books or other publications. Too many, to quote them here! 
A short conclusion is found  in Sect. \ref{concl}.

However, to close this introduction, and put the main section into a proper perspective,  let it be mentioned,  that much  data analysis  work in econophysics  is based on \begin{itemize}
 \item  rank-size, frequency-size concepts,  and usual statistical tests and considerations, 
 \item analysis of the correlation functions of (often the fluctuations in) financial signals
 \item ... often (alas) equal time correlation functions,
 \item though various time scales, and time lags should be much more often considered.   \end{itemize}
 
 Next, economic/financial $models$ are constructed based on developments/generalizations of basic physics models (Brownian motion, Ising and Potts or ferroelectrics models)  modernized into agent based models, often studied on networks. 
 
 Finally,  two basic techniques are used for finding "solutions" or features, i.e.
  \begin{itemize}
  \item  analytic  work, studying one or a set of differential equations,
  \item numerical simulations, often using the Monte-Carlo method.
 \end{itemize}

 Again, let it be emphasized that the note is mainly based  on a few of $my$ contributions, thereby apparently much reducing the scope of the many existing and brilliant investigations. This is not a statement of contempt, on the contrary.  It implies an accent of much modesty.

\section {Micro-econo-physics}\label{Sect2}

There is much data available in financing, banking, option markets, stocks and
currency exchange rates, discount and interest rates. There are  many levels of
observationÊ: individual income, individual expenses,  checking accounts and
savings, public or private accounts, volumes, debts and credits, tellers,
dealers,  bank outlets, businesses,  governments. One is immediately tempted to
play $statistics$, hopefully expecting  to reach some stylized features  of financial matters. 

What should be the number of data points? 
Physicists prone to thermodynamics  and {\it statistical physics} have used from the
very beginning the Avogadro number as an upper  limit. The lower limit is always prone to criticism. Yet, one should remember that there are appropriate tests (Student t-, $\chi^2$, ...). These can be sufficient, when the data has been taken on some kind of {\it equilibrium state}. 

It is worth to point out that there is also no drastic need for considering a large phase space in order to describe complex {\it non-equilibrium systems}. Since Lorenz \cite{Lorenz}, at least, one knows that $three$ coupled non linear differential equations, can lead to the prediction of steady states, cycles, and deterministic chaos, - in a $continuous$ time approximation. At the $discrete$ time level, since Feigenbaum \cite{Feigenbaum80},  
one knows that a  condition, implying the third derivative of the $one-dimensional$  mapping, informs on the possibility of chaotic systems. Thus, there is no need for introducing irrationality or stochasticity, from the first steps, to describe complicated evolutions. Some simplicity is already leading to very complicated features when nonlinearity is implied. Moreover, the phase space (the number of variables)  may always be very  reduced. Of course, economists can always claim that  "this is a too much reduced phase space".  Indeed. What is then the relevant size ? ....

 Clearly, one can always introduce new variables and evolution equations \cite{AsadaOuchi13}.  But one can, thereafter, also  argue that these variables and equations  do  not take into account, e.g.,  (non exhaustive list) the cultural, sport, and "social" aspects of the economic society, - though they are very relevant nowadays!  However, simplicity and intuition are recommended at the starting gate. Results should  be convincing if they recover stylized facts, beside implying forecasting. Nevertheless,  the question of the size of the  relevant phase space in $micro-economy$ problems should be  a relevant study, ... and  some worthwhile information for further  econophysics studies and applications.

Let it be recalled,   at this stage, that beyond equilibrium phenomena,  physicists and mathematicians discovered dissipative structures \cite{dissip}, intermittency  \cite{Vicsek} and  self-organized criticality \cite{soc,socBTW2,bakbook} in non-equilibrium states.   ÊIn fact, cooperative effects like those seen at phase transitions, thermodynamic ones  \cite{PTCPstanleybook,fisher} or geometric  (or probabilistic) ones \cite{percolation}, have been  found to be analogous to  those found, e.g., in financial bubbles, see below.  An essential ingredient, the fractal concepts \cite{Westbooklure,Addison} were already pointed out, "long ago" \cite{peters}. This is one of  the origins of physicist competition to overcome mathematicians who have  invaded economy faculty or arbitration
rooms or brokerage firms. 

Of course, statistical fits and inherent tests are useful \cite{MA4}, including statistical models, but parameters should be controllable and their variation understood. This leads to emphasizing the major difference between "statistician  models", like ARCH and subsequent extensions, and modern econophysics models, based on   "agent" behaviors.  Physicists know the limits of  any model because of inherent hypotheses  and should honestly question the findings. However, the forecasting is better controlled in econophysics than through the many theories which many economists invent and discuss. One famous case is that of financial crashes predictions.

\subsection{ Crashes}  \label{ Crashessubsect}

One spectacular set of features belongs to financial crashes.
 Because of its magnitude, the stock market crash of October 1987 outshines all
downturn ever observed in the past. In one day, the Dow Jones lost 21.6$\%$
and the worst decline reached 45.8$\%$ in Hong Kong \cite{bates}. By comparison, with the most famous
crash of October 1929, the crash was spread over 2 days, a 12.8$\%$ on October 19 followed
by a 11.7$\%$ drop the next day. Note that Lauterbach and Ben-Zion \cite{LBZ93}
found that trading halts and price limits had no impact on the overall decline of
October 1987, but merely smoothed return fluctuations. Another supposedly major
characteristic of the crash of October 1987 is the phenomenon of irresistible
contagion  that the shock arose \cite{Roll89}.

Some application of statistical physics ideas to the description of  such stock market
behavior  was proposed in 
\cite{SornetteAJBouchaudJFI96,Feigenbaum96}. They indicated that  most economic indices 
follow  a  divergence-like power law with a $complex$ exponent,  $m=m '+i\; m"$, see Appendix A.
A rupture occurs at $t = t_c$, the crash time. This law is similar to that of critical points at so-called second order phase transitions \cite{PTCPstanleybook,fisher}, but generalizes the scaleless situation ($m"\equiv 0$) for cases in which discrete scale invariance exists \cite{DSI,DSphysrep}.  Indeed, superposed to the divergence, described by   $m'$, a series of oscillations, with frequency related to   $m"$,  could be observed, - as in the ionic content of water sources before some earthquakes \cite{kobe}, suggesting an analogy including precursors and replicas  \cite{SornetteAJBouchaudJFI96}. Interestingly, the period of these oscillations  is logarithmically reducing as one converges to the rupture point $t_c$, leading to the appellation $log-periodicity$. Alas, the parameter fits are not robust against small perturbations, - due to the numerical instability of a nonlinear seven parameter fit.

 Thus, it has been proposed that the "universal exponent" $m$ is in fact
close to zero  \cite{how,cp}, i.e. the divergence of a financial index $y(t)$ for $t$ close to $t_c$ should rather 
behave like a  logarithmic function,  see Appendix B. In fact, this  is the behavior of  the specific
heat (a "four point correlation function") of the magnetic (2-dimensional) Ising model
\cite{PTCPstanleybook,fisher}. Another type of  similar (so called "essential") singularity in physics is that occurring at the Kosterlitz-Thouless (KT) phase transition \cite{KT72,KT73}, for
dislocation mediated melting. It represents  the transformation from a disordered
vortex fluid state with equal number of vortices with opposite "vorticity" to an
ordered molecular-like state with molecules composed from a pair of vortices with
different polarities across the KT transition.  The logarithmic behavior can also
be observed in the temperature derivative of the electrical resistivity of
magnetic systems at the paramagnetic-ferromagnetic phase transition \cite{Sousa1,Sousa3,Sousa2}.
The behavior is thus generally specific to systems with a $low$  $order$ spatial
dimensionality of the so-called ''order parameter''.

Thus, one can test such a (seven or six parameter) function  on financial indices,  
i.e.,  on one hand there is a
4 or 3-parameter   divergence and on the other hand there is  a log-periodic function, see Appendices A-B, describing the oscillations of the financial index before the crash.

In so doing, one could monitor many indices after 1987. In 1997, the stock market numerical conditions astonishingly looked like the pre-crash period of 1987.  Thus,  in view of  the huge similarity between the 1987 and  1997 behaviors,  
one could come up with  accurately predicting the Oct. 27, 1997 crash \cite{how,cp,TTcrash1,cash1,TTcrash2,cash2}. Much discussion has followed such a prediction, often called "accidental", - by those who did not predict it  \cite{Laloux98nocrash}. However, other successes can be mentioned.   \cite{sornettenikkei,JSL,BBAV00IJTAF,stauijtaf,sornetteNASDAQ,johansen2002comment,bigtokyo,DrozdzPhRep515}. Sometimes, necessarily {\it a posteriori}, one can even detect "near-crashes"; see Appendix B.

 Later on, many investigations have allowed to understand that such a behavior, i.e. a divergence and a log-periodicity of the oscillations,  can arise when the systems present  an endogenous discrete scale invariance, i.e., the system (= market)  has an inner structure which can be {\it  mutatis mutandis} reproduced at different scales (or levels). A sandpile on a fractal-like basis  is a fine analogy \cite{bigtokyo}. It suggests that such a crash is due to a limited number of agents (the most "important/active" ones) who trigger an avalanche. One can therefore consider that there are two basic categories of financial crashes:  endogenous ones and  exogenous ones \cite{DrozdzPhRep515,sornette0210509v1,sh,sornette0412026v1}.  
 
 The exogenous crashes should appear more suddenly because they are  triggered by $quasi$ unexpected news or in physics terms "external fields".   Alas, they are less easily predicted. It seems that they seem to occur more frequently than endogenous crashes.  Moreover, statistical physics studies on triggered phase transitions indicate that  one should expect less universal features in exogenous shocks. On this point,  one should note that a study of sales, when either based on publicity or on  buyer herding, indicates  a different range of relaxation times \cite{sales}: the relaxation time seems to be twice shorter in endogenous shocks than in exogenous ones.  Exogenous and endogenous   after-shocks  sales are thus discriminated by their short-time behavior.  Many studies in econophysics pertain to the recovery of financial indices after a crash, but much has still to be done  in order to  associate such simple analytic laws with the fundamental laws of economy in  a broad
sense.  For some completeness, let the work of  Dro\.{z}d\.{z}  et al. be emphasized \cite{DrozdzPhRep515}, and  their refs. therein.

{\it In fine}, physicist objective ambitions are only to analyze financial data in order to find whether a break point in the series has a precursor, and some after shock  \cite{SornetteAJBouchaudJFI96}. In data analysis as most often done in this matter, the  crash target date represents the ultimate date of a break if the stock market indices continue their extraordinary growth. In brief, the study consists of watching if a major correction is becoming more and more imminent, even if it can take several different forms. Another test point is the crash amplitude.  Finally, the crash duration, the after-shocks and replicas are also   interesting questions.
 
 To be fair, note that   other features should be taken into account when predicting and describing a financial crash. Indeed, financial bubbles and crashes are associated with the phenomenon of {\it volatility clustering} \cite{ContinKirman}, i.e., the existence of successive periods of small and large amplitude of  an index fluctuations.  The volatility clustering is seen through the autocorrelation function of the  index fluctuation amplitudes. Interestingly, one observes    that the autocorrelation function is   decreasing as a  power-law,  like the autocorrelation function of the order parameter, near a critical (thermodynamic) phase transition, e.g.  the susceptibility in magnetic systems \cite{PTCPstanleybook}.  Similarly,   trading volume is positively correlated with market volatility, and both trading volume and volatility show the
same type of Òlong memoryÓ behavior \cite{[36]Lobato}.
 This is understood if considering that the market becomes more and more sensitive to perturbations, as  when  an order parameter is approaching the critical  transition state. An arbitrarily small fluctuation due to some arbitrary agent can trigger
a cascading response of the market  \cite{lux00volatility}. This  also resembles intermittence phenomena \cite{intermitt} and is much reminiscent of $ Self-Organized$ $ Criticality$ (SOC), i.e.,  the tendency of   dissipative systems to spontaneously evolve into a "critical" state \cite{bakbook}.  

 In brief, long memory at crashes, in bubbles,  is directly linked with the effect of volatility clustering, while the log-periodic substructure is due to hierarchical cascades.

\subsection{ Portofolio  selection}  \label{Portofoliosubsect}

Risk must be expected for any reasonable investment. However risk is an intuitive notion that resists formal definition \cite{Holton}.  One expects that a portofolio should be constructed such as to minimize the investment risk in presence of somewhat unknown  fluctuation distributions of the various asset prices in view of  obtaining the highest possible returns \cite{Markowitz,Markowitzbook,Marshall}. Usually analysts recommend investment strategies based e.g. on Òmoving averagesÓ, Òmomentum indicatorsÓ. Another sort of data analysis technique, can be adapted to portofolio management \cite{bronlet3Tokyo}, leading to forecasting and prediction, with some "risk": it is known as the Zipf technique,   originating in work exploring the statistical nature of languages \cite{z1}; see Appendix C.
 The Zipf method previously
   developed as an investment strategy
   (on usual financial indices) \cite{bronlet1,bronlet2} can be adapted
   to portofolio management.
   
 In brief, if a crash can sometimes  be predicted as suggested here above,  the next fundamental step is to predict the amplitude of the fluctuations before     (but also after) the event.  Indeed, this information should allow economic agents to hedge portofolios on derivative markets through call or put options. The problem reduces in fact to extract the true background contribution  of an index.  Indeed,  it  represents the `"natural"  (noisy and structural) evolution of stock indices out of any euphoric ("anomalous") bubble evolution. This is a classical  question in thermodynamic phase transitions before searching for critical exponents characterizing power law behaviors. This has been somewhat tackled through Zipf  "lazy method" of mere counting $up$ or $down$ fluctuations, and expecting some "only smooth non-equilibrium" of the market. 
 
 The index time series of the closing price of some stock is  translated  
 into a series of  letters, e.g. taken from a two character ($u,d$) alphabet, corresponding to $up$ or $down$ fluctuations.  (This can be generalized, of course). One can next search for all possible  words of $m$ letters, and investigate 
the occurrence of such words. A statistical table can be built, for the probability of findings such words. Some investment strategy can then be decided, assuming the probability to vary weakly with time.

  In \cite{bronlet3Tokyo},  e.g.,  two portofolios were invented, based on stocks in 
  the $DJIA-30$ and the $NASDAQ-100$.  After some time, two strategies with  different weights for the shares in the portofolio  at buying or selling time, were imagined.   For the next few years, the  yearly expected return, variance,
Sharpe  ratio and $\beta$  were calculated in each case; see Appendix D  for the relevant definitions and comments. The best returns and
weakest risks mainly depend on the  chosen word length inducing the strategy. Even though some risk values could  be large,  returns were sometimes very high \cite{bronlet3Tokyo}. Note that the investment time included exogenous and endogenous events. Much variability is clearly possible, depending on investor's choices.

One disadvantage of the Zipf-method is that it  does not easily distinguish
 persistent and antipersistent sequences. Another  method, a "detrended fluctuation analysis method",  (DFA)  \cite{s25PRE49}  can also be used to build investment strategies. It is described here when discussing econophysics work on the foreign currency exchange markets.

\subsection{ Foreign Exchange Markets}  \label{FXsubsect}

The technical details of  "detrended fluctuation analysis method",  (DFA)  are left for Appendix E.  Let it be known that the technique originated in  Stanley
group \cite{s25PRE49} for sorting out $DNA$ coding and noncoding regions.  First note that, like for crashes, the (daily and weekly seasonal) volatility is of interest  in the foreign exchange market \cite{Dacorognaetal}. In particular, it was  
shown \cite{VA26} that the time dependent signal for  foreign exchange
currency rates does $not$ obey the Brownian motion rule (coin tossing), but is
rather a $fractional$ Brownian motion \cite{Addison,Westbooklure}. The "diffusion"  of the particle (=   share price or index value) does not
evolve like the square root of time,  $t^{1/2}$, for large $t$, but has an exponent quite different from
0.5. Coherent and random sequences  can be  clearly  seen  in the  financial fluctuations 
\cite{VA26,SRFXM1,SRFXM2}. By the way, the analysis is robust against various {\it a priori} trends \cite{MA299,DFA1}.

As an application, mention the evolution of  foreign exchange currency rates, both for emerging markets and well controlled ones.   The  exchange rates seem to belong to different ''categories'' depending on the involved currenciesÊ: strictly
regulated european, hard dollar zone, emerging markets, etc.  \cite{Ki2,KIMAfalseEUNikkei}.      The parameters are similar   to those known in turbulence  \cite{intermitt}. 

A multi-affine fractal-like analysis \cite{mufract}, which however for lack of space will not be discussed further here, has also been implemented; see also  \cite{INDIANBOOKch9} and Appendix E. 

Thus, not only long term correlations exist and can be implemented for predictability, but some systematics of the short range correlation functions through low order $i-variability$ diagrams \cite{Babloyantz} for short range correlation evidence in,  e.g.,  exchange rate (and Gold price),   are  also observed  \cite{Ki1}.

Very interestingly the features, i.e. like the  characteristic exponent describing the evolution, present sometimes remarkable drastic turns. The features can be  daringly associated to $economic$  events  following "$political$" events or to ome panic storm spreading over financial markets.  It turns out that exogenous causes, e.g., the Gulf War, the technological bubble explosion, ..., are  {\it a posteriori } (of course),  well seen \cite{VA26}.  One can also distinguish   an exponent  behavior   quantitatively  modified  after some policy move for better control and in order to avoid panic or to heal a
crisis. However, one can also observe/guess that the  market agents thereafter likely search  for the best subtle loopholes in the regulation, in order to avoid the most severe constraints, then slide off the policy main stream   \cite{VA26}., ....  before new rules are
decided upon.

\subsection{ Predictability} \label{Predictabilitysubsect}

 Thus the most important question follows Ê: knowing a  distribution law of market fluctuations, may a winning strategy may be thought of and is predictability possible, on financial indices, on the currency exchange market, on the option market,  ... ?  
 
 Yes. Among many cases, for example, several exchange rates and futures were studied in this respect, on a 16 year period. A robust (DFA) law for
persistent and anti-persistent fluctuations was obtained. Thus, one was virtually
able to increase an input capital by a factor of 40, but with less success for the
price of gold, under ''perfect'' conditions, i.e. no tax no  broker fee. 

Usually, economists  (i) do not admit such a type of predictability because they object about the lack of
psychological features plus various theoretical matters, and (ii) estimate
that the market is  justified by real earnings of actors, in the long run. In the
short run, economists of course admit  fluctuations, based on (sudden, or not)     events which
create an atmosphere of optimism or pessimism that makes market actors
systematically over- or underestimate future earnings. Economists claim that
physical models cannot predict such surprises that suddenly change the
optimism/pessimism in the market. They are right, except that physicists do not
have to make such models. No physicist can predict an electrical power failure in
the laboratory. But if a source of heat is removed, it is known that water may freeze when the temperature drops below 0$^{\circ}$C, - that is well known to physicists.

\subsection { Price and asset distribution }\label{Pricesubsect}

The word "price" has been mentioned several times here above. One more comment is in order, to show a connection from basic physics to finance through econophysics, on this matter.

Many recent observations have indicated that the traditional
{\it equilibrium market hypothesis} ($EMH$; also known as Efficient Market
Hypothesis) is unrealistic.  For example,  long-term memory  effects are in stock market prices \cite{[37]Lo} and volumes \cite{[36]Lobato}), imply a non-gaussian distribution of these.   In fact,  a price evolution along the EMH is analogous to  a Boltzmann
equation in physics, thus having some bad properties of mean-field approximations
like a Gaussian distribution of price fluctuations  \cite{MA359kinetic}, - since such distributions and related ones have "fat tails" \cite{BBM63,BBM67,[37]Lo,[36]Lobato}. This {\it kinetic theory for prices} can be simply derived, considering in a first approach that market actors
have all identical relaxation times, and solved within a Chapman-Enskog like
formalism. In so doing,   an equation of state is obtained linking  a
$pressure$, a $temperature$, i.e. the inverse of the relaxation time,  and a $volume$, - the price ({\it being
taken as the order parameter}) of a stock. This may lead to further studies  on relations between intensive and extensive variables \cite{PTCPstanleybook} with applications  as in "less simple" technical analysis schemes \cite{epjb_gta,MAKI433}.

A Boltzmann-type master equation for the  problem  of asset exchange  and an analytic solution have been derived  in \cite{PRE72.05.026126Chatt}. It was shown also by  Chatterjee, Chakrabarti and Manna (CCM) that kinetic models of money exchange with the added feature of `savings' can nevertheless produce self-organization and a heavy-tailed money (asset) distribution for traders with distributed savings \cite{CCM}.  In   \cite{CCM}  a "Pareto region"  is found in the probability distribution, with exponent value $\simeq$  1. Previously,  Chakraborti, and Chakrabarti,  \cite{Eur. Phys J B 17 (2000) 167} had numerically
studied a random exchange model, which conserves money, and in which each agent saves a fixed fraction of their instantaneous money.   This model \cite{Eur. Phys J B 17 (2000) 167} was later reanalysed   by Das and Yarlagadda \cite{DasYarlagadda} using a Boltzmann transport type formulation. 

Instead of savings one can investigate taxes. A very simple model of a closed marked first with tax-free exchange of goods and wealth  has been studied in \cite{MAAPwealthPhA373}: the amount of goods and the accepted price, as well as the trade decisions have no relation to trends, previous activity, anticipation, savings, etc. Also partners for trading are chosen randomly among all agents. Such a free market stabilizes itself "rapidly", - the average price, number of transactions, number of
goods sold and money paid for them, become asymptotically fixed in time. In some sense there is an
intrinsically generated (ÔÔself-organizedÕÕ) utility function. The distribution of money and goods shows a
stratification of the ÔÔsocietyÕÕ. (This was further confirmed in a study in which there is competition between  size-dependent peer agents \cite{Caram}). There seems to be a trend
away from Pareto's values of 1.5 to slightly higher values. It can be wondered  if this represents the impact
of socialism policy since a high value "fat tail"  is associated with greater redistribution of money, as   found \cite{MAAPwealthPhA373}, whereas a low value of the Pareto exponent suggests that the rich really are rich and the poor are poor.  When part of the money disappears from the market under the form of
ÔÔtaxesÕÕ, at each transaction, characteristics are quite similar to the ÔÔno taxÕÕ case, though
there is a difference  in the distribution of wealth, i.e., the poor gets poorer and the rich gets
richer.

\section { Macro-econo-physics Aspects}  \label{Macroecono}

Rather more recently econophysics  has been enlarged to  studies of $macroeconomy$.  It arose in the case of \cite{ACP1} when searching to confirm or infirm, within  econophysics modeling,  political statements, like "The North must help the South! It's good for the North economy".   However,  this is is hard to swallow by workers who see their work being delocalized. When will the growth part of the business cycle come back? 

The  ACP model \cite{ACP1,ACP3} has  many ingredients : different  geographic regions,  initial concentration(s), economic field time sequence(s), selection pressure, diffusion process rule(s), enterprise-enterprise ÒinteractionÓ(s), business plan(s), number of regions, enterprise evolution law(s), and
economy policy time delay implementation,  all presupposed to be known for the
Monte-Carlo simulation. It is found that the model even in its simplest forms can
lead to a large variety of situations, including: stationary solutions and cycles, but
also chaotic behavior.
 
This complexity led to simplify the macro-economic question  to read :{\it is there any economy globalization?"}. 
The globalization process \cite{britanica}  can be discussed from a
political \cite{political}, cultural \cite{cultural}, scientific co-operation \cite{co-operation1} or economy \cite{economy} point of view. The problem becomes so fashionable nowadays that even  so called antiglobalization movements and meetings are organised \cite{antiglob,antiglob1}.  In   \cite{JMMA486,JMMA545},  the definition  was restricted such that one could study  the case of  the world economy in  a quantitative aspect, i.e.
\begin{itemize} \item
A globalization process in economy is understood as the increase of similarities in development patterns.
\end{itemize}

 The most important questions which arises thereafter this definition pertains on how to measure { \it similarities in patterns} and which economy parameters have to be taken into consideration for doing so.  Much debate exists, - many admit that the different approaches might intrinsically imply different  (types of)  conclusions.

 In \cite{JMMA486,JMMA545}, "similarities" were searched for by measuring { \it distances} between  the Gross Domestic Product (GDP)  time series. The  GDP seems to be the  most representative parameter describing the status of an economy, since, {\it mutatis mutandis}, it is defined for all countries.   Data was taken from 
 $ http://www.ggdc.net,2006$ \cite{ggdc}.   {\it Differently defined "distances"}  between pairs of GDP series were considered, either based on the linear cross-correlation coefficient, i.e. a {\it statistical} distance,  or  on a measure of  the disorder level in the GDP evolution, i.e. an {\it entropy} or information distance.   It was found that the  time averaging of the distances over finite size time windows is  fundamental. It implies considerations, on policy effects, due to time delay  in implementing policies.  Nevertheless,  some country  hierarchy is  obtained.  Whence network structures  can be constructed based on the
hierarchy of distances. It  was shown that the mean distance between the most developed
countries on several networks   decreased in time. 

 By the way, it was  shown that the entropy distance measure is more suitable in detecting a globalization process than the usual statistical (correlation based) measure. Note that other  macroeconomic indices, similarly studied in  \cite{JMMA562}, confirm the findings. Moreover, the results indicated that  the EUR introduction and the Maastricht agreement constraints induced  the start of a deglobalization between European countries!.

See also, but using a different measurement technique \cite{PhA388.09.3527GDP LA}, a related study on  hierarchies and structures in the Gross Domestic Product per capita
fluctuation in Latin American countries

In the same line of questioning, it might be audacious  to consider a country hierarchy when describing them through network schemes   based on some  ranking  through macroeconomic indices. However, not only in doing so one obtains a way to elaborate  on structured clusters, but  also connect to Hamiltonian mechanics. More  considerations can be found in  \cite{gligor512,gligor537,gligorAOI}.

\section { Perspectives \& Prospects}  \label{discuss}

Scientific progress is expected to depend on the combination of experimental and
theoretical investigations that provide deep understanding of   phenomena,  with an absolute need  for   a high reliability of results. Moreover, simulation experiments can nowadays be part of theoretical or experimental advances; see  the case of "sociophysics" in \cite{DS468,DS502}. The simple ideas  proposed here above on  thermodynamic, and sometimes geometric, 
phase transitions  lead to financial markets studies and understandings, including sociological aspects.
All that seems to be derived from the study of a dynamical state propagation in a random medium, - lattices or networks. The
fundamental ideas refer to   correlations in fluctuations \cite{fisher},  invasion percolation \cite{percolation}, and self-organized criticality \cite{soc,socBTW2}. 
   
Many, many, other topics could be presented. For lack of space,  let us point to a very (very) small set, with one reference only, for the interested reader to start from:
   
   \begin{itemize}
   \item increasing returns and economic geography; see Krugman \cite{krugman1990increasing},
   \item modelling and pricing  derivatives, even weather; see Alaton et al. \cite{alaton2002modelling},
   \item  discussing inequalities of wealth distribution, with the Gini coefficient; see Pianegonda  and Iglesias \cite{pianegonda2004inequalities},
  \item assessing information flow time delay in the formation of economic cycles;   see  Miskiewicz and  Ausloos,  \cite{476JM} 
   \item  controllling shareholding networks through cross and joint ownership; see  Rotundo and D'Arcangelis \cite{rotundo2010ownership},
   \item  finances in  sport; see Torgler \cite{Torgler},
   \item price auctions,  in e.g. art   or  other cultural matters; see Reddy and Dass \cite{ReddyDass}.
  \end{itemize}

{\it In fine}, it is time to comment on "methods" and "models", -   in fact adding to the remarks at the end of the Introduction Section.
 
\begin{itemize}\item Methods \end{itemize}

Tools of time series analysis, varying the scale of resolution, are very useful to characterize important features of complex systems. It is strongly recommended to Fourier transform whatever signal is investigated \cite{MA378,MA410}, in order to sort out the most important periodicities in the system: yearly, seasonal,  weekly, daily, ... or others. This leads also to measure a parameter, the phase, which may depend on  e.g.  different fiscal year terms.  

Beside the techniques not mentioned so far, let the Recurrence Plot Analysis  be pointed out. It has served to study critical regimes, like crashes \cite{AFMARCARQA05,AFMARCARQA06,KGBBRCRCA10}, among other features.
The Theil, entropy-like mapping, of a time series  \cite{JMTheil} is also very interesting, in particular when  searching  whether Benford law \cite{Newcomb,Benford}  is valid \cite{PCMATheilBenford}.

\begin{itemize}\item Modeling \end{itemize}

 Let it be re-emphasized  that several scientific issues can be found: e.g. (i) creating models based on financial insights and  mathematical principles, (ii) calibrating models based on market information, and (iii) simulating  models using specific algorithms. 

From   the statistical  analysis of "size-frequency distribution",   an attachment process is  often seen (through the "fat tails" \cite{BBM63,BBM67})  as the primary cause of the distribution (of returns, e.g.) evolution \cite{ContBouchaudherd}.  Such constraint aspects must be  introduced in network descriptions of financial and economic matters. However, the modern description of human societies through networks  is  often lacking other mandatory ingredients, i.e. the non scalar nature of  the nodes, and the non binary aspects of nodes and links, though for the latter this is already often taken into account, including directions, - but this quite complicates the algebra \cite{GRMA_ACS}.

Coming back to roughly the two main  themes (which might be thought as approximations) in micro-econo-physics  :  
 there are based on the supposedly 
different goals and strategies of "players" (in other words "traders" or more generally "agents"): chartists or fundamentalists, in so calling,  remaining within the distinction made by micro-economy theories.   

 One has to consider competition like processes.  
 Future trends in econophysics studies should connect better to reality in tying such features of agents and society.  
Indeed, changes  in economic behavior can occur either due to ''heterogeneous agent interaction'' processes or due to ''external field'' constraints, - or both.  

Beside the basic physics models recalled in Sect. I, and for which there are numerous references, let the recent fluctuating mass Brownian particle \cite{MARLPRE73.06} be understood as a physical analogy to  a "price.volume" variable \cite{epjb_gta,MAKI433}, - what is often  in fact   the real constraint of investors.

\section { Conclusions}  \label{concl}

One  should re-emphasize that the very
relevant item, which much pleases "fractalists"  and statistical physicists is the
discovery (after several attempts) that the markets are not really ''efficient'', but is "scaling" \cite{MS24}, -  the (truncated) Levy
distribution was rather a universal one  in order to describe fluctuation correlations,  and evolving like order parameters in physics \cite{BouchaudCont}. 
Bachelier analysis and the Gaussian hypothesis are not true \cite{peters}.  The Black-Scholes hypotheses are incorrect  \cite{BouchaudBlackScholes94}.   Discrete Scale Invariance \cite{DSphysrep}  has to be inserted from the start in network models.

Economists scorn physics models because the former ones claim to have also
endeavored the prediction of short term fluctuations  by examining past
fluctuations (chart analysis) or approximating stock prices by wave structures
(Elliott waves). These models, obviously, can only be a source of profits if they turn out to be a self-fulfilling prophecy. This can happen if many market actors believe in
such theories. For example indeed,   it has been shown that the mobile average
technique is wholly unrealistic for predictability purposes \cite{mobav}.
Therefore, economists are sending warnings about their own techniques, though not
seeing differences with respect to physicist approaches. As a famous 'non-econophysicist' said once: {\it If "it" occurs once and I can explain "it", I can explain "it" many times and "it" can occur many times}.

The ÒneedÓ for Òall thatÓ can be criticized by physicists, mathematicians and
economists, from first principles or on mathematical grounds. The econophysics approach
should be taken with caution, indeed. The robustness and soundness of models are
fundamental questions. Models should be predictive, and should be tested. Zhang
\cite{Zh1} claims that there are too many directions of investigations.  This
could deserve the research. I believe that they are too many topics available at
this time for the scientists involved as well. However, some self-organized
framework might emerge.  I am on this point rather optimistic \cite{MAinEN}.

   \bigskip  
{\large \bf Acknowledgements} 
\bigskip  

 Thanks to P. Clippe, A. P\c{e}kalski,  and D. Stauffer,  for numerous discussions and friendly critical comments.  
 This contribution  is   part of scientific activities in (i) COST
Action TD1210, "Analyzing the dynamics of information and knowledge landscapes - KNOWeSCAPE", 
and  (ii) COST Action IS1104, "The EU in the new complex geography of economic systems: models, tools and policy evaluation".

    \bigskip  

{\large \bf Appendix A : Log-Periodicity and divergence} \label{App_logPdivm}

In two independent works
\cite{SornetteAJBouchaudJFI96,Feigenbaum96}, it was indicated that  most economic indices, $y(t)$,
follows a power law with a complex exponent,  $m=m '+i\; m"$, such that
 \begin{equation} \label{crashpldiverg}
 y(t) = A + B {\left( \frac{t_c-t }{ t_c} \right)}^{-m'} \left[ 1 + C \left(
 \omega \ln{\left( \frac{t_c-t }{ t_c} \right)} + \phi \right) \right]  \end{equation} 
 for  $t  <  t_c$,
 where $t_c$ is the crash-time or rupture point, the other symbols being parameters.   
A divergence occurs at $t = t_c$. This law  generalizes the scaleless situation at  thermodynamic critical points, so-called second order phase transitions \cite{PTCPstanleybook,fisher}, to cases in which discrete scale invariance exists \cite{DSI,DSphysrep}.  Superposed to the divergence, a series of oscillations could be observed, - as in the ionic content of water sources before some earthquakes \cite{kobe}. Interestingly, the period of these oscillations converges to the rupture point $t_c$. The modelization leads to a seven parameter function for fitting the data. 

  Various values of $m'$ were reported ranging from 0.53 to 0.06 for various indices and events (upsurges and crashes) \cite{Feigenbaum96}, while $m'$ ranges from  0.7 to   0.33 in  \cite{SornetteAJBouchaudJFI96}.  This lack of "universality" suggests a strong error bar effects, due to many possible causes, thereby suggesting some simplification of the function as
 \begin{equation} \label{crashlogdiverg}
 y(t) = A + B \ln{\left( \frac{t_c-t }{ t_c} \right)} \left[ 1 + C \left( \omega
 \ln{\left( \frac{t_c-t }{ t_c} \right)} + \phi \right) \right] \hskip 0.7cm
 \end{equation} for   $t  <  t_c$. Moreover, as mentioned in the main text, this functional form reproduces known intriguing phenomena in physics, several of them allowing to mimic financial phenomena by analogy.
 
 Whatever the functional form of the divergence, the log-periodic structure has been found in many studies. Beside those in the main text, let us mention \cite{drodz,viz,ALS}.

   \bigskip  
{\large \bf  Appendix B: Crash predictability technique} \label{App_logPdivlog}
 
 Thus, one can test such a (six parameter) function) on financial indices,  
i.e.,  on one hand there is a
3-parameter power law divergence (fixing $m=0$), and on the other hand there is  a log-periodic function, see Appendix B, describing the oscillations of the financial index before the crash.  

In so doing, one has monitored many indices after 1987. In 1997, the stock market numerical conditions astonishingly looked like the pre-crash period of 1987.  Thus,  in view of  the huge similarity between the behaviors in 1987 and in 1997, and admitting some universality of behavior, one could come up with  accurately predicting the Oct. 27, 1997 crash \cite{TTcrash1,cash1,TTcrash2,cash2,how}. 

Practically, one can dissociate the fit to the divergence  from the fit to the oscillations, whence having two 3-parameter fits. One obtains $two$, usually different, $t_c$ values, i.e. $t_c^{(d)}$ and $t_c^{(o)}$. Note that these values of $t_c$ evolve  with time during the signal analysis. If they converge to a single $t_c$, one can predict the theoretical crash time. The distance  $t_c^{(d)}\;-\;t_c^{(o)}$ may also evolve smoothly,  - sometimes indicating a  "near-to-crash"  or a  "missed crash". A "crash risk" notion can be invented.

   \bigskip  
{\large \bf Appendix C: Zipf method} \label{App_zipfmethd}

 The Zipf method \cite{z1}  consists in counting the  frequency of "something" in view of some ranking. In the original case,  it was the number of words in a text.     
 Zipf \cite{z1}  observed  that a large number 
of  such distributions, $N_r$ can  be approximated by a simple  {\it 
scaling (power) law} $N_r = N_1/r $,   where  $r$ is the ranking parameter, with 
 $N_{r } \ge N_{r+1}$, (and obviously  $r<r+1$). 
 
 More generally,  one can translate a time series into a text by choosing an appropriate alphabet to mimic the signal variations.  For example, in finance, one can analyze in such a way, the number of fluctuations of a given size in a signal.  One choose a two letter alphabet ($u,d$) for $up$ and $down$ fluctuations; or a 3-letter alphabet ($u,s,d$) where $s$ corresponds to $small$ (as chosen by the analyst) fluctuations; or a five letter alphabet ($u,p,s,n,d$), etc.  One counts the  "words", overlapping or not, either with  equal  "length", or not. One ranks next the "events" according to their frequency,  in frequency decreasing order; the most frequent getting the rank $r=1$, etc.  
The $rank-frequency$ relationship,  i.e. the  frequency  $f$  of the  occurrence of an "event" relative to  its rank $r$, usually reads like an inverse power law, $f\sim r^{-\zeta}$. 

Note that one can also ask \cite{Pareto}  how many times one finds 
  an  "event" greater than some size $x$, i.e. the "size-frequency" relationship. Pareto's  found out that  the cumulative distribution function  of such events, i.e. the
number of events larger than $F$, (also) follows an inverse power of $f$, i.e. $P\;[f>F] \sim f^{-\lambda}$ \cite{Pareto}. Theoretical work leads to $(1/\lambda)+  \zeta=2$. Recall that Pareto discovered that  income distribution does not behave in a Gaussian way, but exhibits ``heavy tails".

  \bigskip  
{\large \bf Appendix D:  Technical definitions for risk evaluation} \label{App_Sharpe}

 Call the relevant  financial index, so called market ($M$), and portofolio ($P$) variance   $\sigma^{2}_{M}$ and $\sigma^{2}_P$, respectively.  The positive square root is called the standard deviation.  Let $E(r_{P})$ be the yearly returns.
   
    The Sharpe ratio $SR$,   given by $SR$ = $E(r_{P})$ / $\sigma_P$,
   is considered to measure the portofolio
   $performanc$e \cite{Sharpe}.
   
   The $\beta$  is given by $cov(r_P,r_M)/ \sigma^{2}_{M}$
   where the $P$ covariance $cov(r_P,r_M)$ is measured with respect
   to the  relevant  financial index,   i.e. $cov(r_P,r_M)= E(r_P,r_M)-E(r_{P})\;E(r_{M})$. 
   The $\beta$  is considered to measure the portofolio $risk$.
    
       \bigskip  
{\large \bf   Appendix F:  Detrended Fluctuation Analysis (DFA)} \label{App_DFA}

 The Detrended Fluctuation Analysis (DFA) technique was introduced  in order to investigate long-range power-law correlations along
DNA sequences [11, 12]. The method consists in dividing the whole discrete
data sequence  
$X=\left\{ x_{i},i=1,...,N\right\} $ (in finance,  $X$ is  a time series) of length $N$ into  non-overlapping boxes, each containing
$n$ points.  It is suggested that the  "first" box contains the "last" data points.  In so doing, when $N$ is incommensurate with a positive number $k$,  one leaves out   the "oldest"  signal values. 

 In the k$^{th}$ box, a "new" signal  is calculated: the wholly reduced-cumulative
sum $Y_{i}^{(k)}$ 
  \begin{equation}
 Y_{i}^{(k)}(n)=\sum_{j=1}^{i}\left(x_{j}-\left\langle x\right\rangle _{N}\right) 
\hspace{10mm}  i=kn+1-n,...,kn
\end{equation}
 and $\left\langle x\right\rangle_{N}=(1/N)\sum_{j=1}^{N}x_{j}.$ 

   In each box,  one defines the local trend
   \begin{equation}\label{local}
z^{(k)}(n) = a^{(k)}\;n + b^{(k)}
   \end{equation}
 as  the ordinate of a linear least-square fit of the data points, in $that$ box. One should remark that a trend $z(n)$ different from
a  first-degree polynomial can also be used, such as the cubic trend  \cite{MA299}. 

Let 
     \begin{equation}\label{F2}
F_k^2 (n)=\frac{1}{n}\sum_{i= kn+1-n}^{kn}{\left(Y_{i}^{(k)}(n)-z^{(k)}(n) \right)^2},
   \end{equation}
   The detrended  fluctuation function   is then calculated by
   averaging $F_k^2 (n)$ over all the  equal size intervals ($ k= 1,2,...,\left(\frac{N}{n}-1\right).  $)   This gives a function depending on the box size $n$: $ <F^2_k(n)>  $. The  calculation is repeated for  all possible different box sizes $n$. 
   
   If the $X$ data values  are randomly uncorrelated variables or short range correlated variables, the behavior of $\sqrt{\left<F^2_k(n)\right>}$  is expected to be a power law 
      \begin{equation}\label{F2alpha}
f(n)=\sqrt{\left<F^2_k(n)\right>} \sim n^{\alpha},
   \end{equation}
   where $\alpha$ is in fact nothing else that 
 the so-called Hurst exponent  $H$ for fractional
Brownian motion \cite{Addison,Westbooklure}.
It can be useful to recall  \cite{Addison,Westbooklure} that the power spectrum of such stochastic  signals is characterized by a power law with an exponent  $\beta =2\alpha-1 $, which itself can be related to the fractal dimension of the signal.

The cases $\alpha > $  1/2 and $\alpha  < $     1/2 should be physically distinguished
 For   $\alpha >$ 1/2 there is  so called persistence, i.e. $ C  >  0$.
 The $\alpha = 0$ situation corresponds to the so-called white noise. 
For   $\alpha <$ 1/2 the signal is said to be antipersistent, thus  apparently very "rough". 
 
   For complementing Appendix E, the autocorrelation function   $C(i,t)$ of the signal
$\{X_i\}_{i=1}^{N}$ is defined as 
\begin{equation}
C(i,t)=\frac{\left<\left(X_t-\mu_t\right)\cdot
\left(X_{i+t}-\mu_{i+t}\right)\right>}{\sigma_t\cdot\sigma_{i+t}}   
= 2^{2 \alpha -1}-1,
\end{equation}
where
$\sigma_t$ is the standard deviation and $\mu_t$ is the mean value  at
the moment $t$.

The cases $\alpha > $  1/2 and $\alpha  < $     1/2 should be physically distinguished
 For   $\alpha >$ 1/2 there is  so called persistence, i.e. $ C  >  0$.
 The $\alpha = 0$ situation corresponds to the so-called white noise. 
For   $\alpha <$ 1/2 the signal is said to be antipersistent, thus  apparently very "rough".

 \end{document}